\documentclass[letterpaper,english,preprint, aps,nofootinbib]{revtex4-1}
\usepackage[T1]{fontenc}
\usepackage[latin9]{inputenc}
\usepackage{babel}
\usepackage{tipa}
\usepackage{tipx}
\usepackage{amsmath}
\usepackage{amssymb}
\usepackage[unicode=true,pdfusetitle,
 bookmarks=true,bookmarksnumbered=false,bookmarksopen=false,
 breaklinks=false,pdfborder={0 0 1},backref=false,colorlinks=false]
 {hyperref}

\makeatletter

\pdfpageheight\paperheight
\pdfpagewidth\paperwidth

\makeatother

\begin{document}
\title{Quantum Chaos and Unitary Black Hole Evaporation}
\author{David A. Lowe}
\email{lowe@brown.edu}

\affiliation{Physics Department, Brown University, Providence, RI 02912, USA.}
\author{L\textipa{\' a}rus Thorlacius}
\email{lth@hi.is}

\affiliation{Science Institute, University of Iceland, Dunhaga 3, 107 Reykjavík,
Iceland}
\affiliation{Nordita, Stockholm University and KTH Royal Institute of Technology,
Hannes Alfvéns väg 12, SE-106 91 Stockholm, Sweden}
\begin{abstract}
The formation and evaporation of small AdS black holes in a theory
with a holographic dual is governed by the usual rules of quantum
mechanics. The eigenstate thermalization hypothesis explains the validity
of semiclassical gravity for local bulk observables and can be used
to quantify the magnitude of quantum corrections to the semiclassical
approximation. The holographic dual produces a basis of black hole
states with finite energy width, and observables that are smooth functions
on the classical phase space will self-average over a large number
of energy eigenstates, exponential in the Bekenstein-Hawking entropy
$S$, leading to results that are consistent with semiclassical gravity
up to small corrections of order $e^{-S/2}$. As expected, the semiclassical
description breaks down for transition amplitudes that contribute
to the unitary S matrix of the holographic theory.
\end{abstract}
\maketitle

\section{Introduction\label{sec:Introduction}}

We revisit the black hole information problem \citep{Hawking:1976ra}
in the context of the AdS/CFT correspondence \citep{Maldacena:1997re,Gubser:1998bc,Witten:1998qj}.
In this setting, the dual gauge theory provides a unitary time evolution
by construction but the implementation of unitarity on the bulk gravitational
side is less clear. Much of the work involving black holes in AdS/CFT
has focused on ones that are large compared to the AdS length scale
and well described using finite temperature quantum field theory \citep{Witten:1998zw}.
However, the standard formulation of the information problem involves
the formation and subsequent evaporation of a macroscopic black hole,
\textit{i.e.} large compared to the Planck scale, in asymptotically
flat spacetime. In an asymptotically AdS background this process is
most closely modeled by considering a macroscopic black hole that
is small compared to the characteristic AdS length scale \citep{Lowe:1999pk}.
The black hole is taken to be macroscopic so that it carries a significant
amount of information and evaporates slowly, but it is taken to be
small enough for the black hole lifetime to be short compared to the
AdS light crossing time to avoid issues arising from the global causal
structure of AdS spacetime. This requires a large but finite separation
between the AdS scale and the Planck scale on the bulk side of the
duality.

One goal of the present work is to extend holographic methods to the
study of small AdS black holes by combining Mukhanov's approach to
counting black hole states \citep{mukhanov} and the HKLL construction
\citep{Hamilton:2005ju,Hamilton:2006az} for mapping small perturbations
in the asymptotic region of the spacetime into the dual gauge theory.
A similar construction for asymptotically flat spacetime was recently
given in \citep{Liu:2021tif,Liu:2021xlm} based on the conformal symmetry
of the celestial sphere \citep{Pasterski:2016qvg}. Further progress
relating asymptotically flat holography to the vanishing cosmological
constant limit of AdS/CFT can be found in \citep{Lowe:2020qan} and
it is straightforward to extend our results to that setting. With
a holographic map for small AdS black holes in place, we can apply
the rules of quantum mechanics and methods from quantum chaos theory
to estimate the size of corrections to the semiclassical approximation
for observables outside the black hole as well as for observables
measured by infalling observers entering the black hole. 

In order to keep the presentation as simple as possible, we will mainly
focus on small AdS black holes in 3+1 dimensional spacetime but our
arguments can be extended to general dimensions. In fact, depending
on the precise holographic setup, the small black hole Schwarzschild
radius may also be small compared to the characteristic size of the
geometry transverse to AdS in a consistent string or M theory background
and in this case the favored configuration is a higher-dimensional
black hole that is localized on the transverse compact space \citep{Gregory:1993vy}.
Fortunately, the two main ingredients in our construction,\textit{
i.e.} Mukhanov's black hole counting and the HKLL formalism, are both
easily adapted to general dimensions and our arguments go through
in this case as well.

The paper is organized as follows. In Section \ref{sec:Black-hole-evolution}
we briefly review Mukhanov's approach to counting black hole states
and then outline the construction of finite-width wave packets formed
out of quasi-local bulk operators that satisfy the spatial and momentum
constraints of Mukhanov states. In Section \ref{sec:Quantum-Chaos}
we turn our attention to the formation and evaporation of small AdS
black holes from a holographic perspective. A key observation is that
due to the finite black hole lifetime the gauge theory state that
corresponds to an isolated black hole has a natural energy width and
necessarily involves a superposition of a large number of energy eigenstates.
The energy eigenstates themselves do not correspond to smooth semiclassical
geometries but a semiclassical description for appropriately defined
black hole observables emerges via eigenstate thermalization. In Section
\ref{sec:Infalling-Observables} we focus on observables that can
be measured by infalling observers in a laboratory that enters the
black hole. Local bulk operators can be constructed and evolved across
the horizon along timelike geodesics and we argue that their expectation
values can be computed to a good approximation by semiclassical methods
until they approach the region of strong curvature near the singularity.
Section \ref{sec:Conclusions} contains some concluding remarks.

\section{Holographic Black Hole Evolution\label{sec:Black-hole-evolution}}

Consider the formation of an isolated small AdS black hole in a theory
with a holographic dual. We would like to identify a set of semiclassical
bulk initial states that can account for the black hole entropy and
map them to states of the dual field theory. Here we have in mind
realizing these states within a single CFT dual to a single asymptotically
AdS region. The bulk states will evolve forward in a semiclassical
geometric theory while the corresponding field theory states undergo
unitary time evolution. By evaluating expectation values of observables
we can estimate the size of the corrections to the semiclassical approximation
that are imposed by unitarity. 

As was observed by Mukhanov \citep{mukhanov}, the overwhelming majority
of initial states in the bulk that contribute to the black hole entropy
will correspond to the time-reverse of outgoing states produced by
the evaporation of a black hole of the same mass, \textit{i.e.} states
consisting of a collapsing cloud of out-of-equilibrium radiation directed
toward the location where the black hole forms over a time period
of order the black hole lifetime. The set of possible initial states
of course also includes low-entropy configurations like incoming spherical
shells of matter or stellar interiors undergoing gravitational collapse
but the dominant contribution to the black hole entropy comes from
incoming radiation states of the form considered by Mukhanov.

To enumerate the incoming Mukhanov states, we find it more convenient
to consider the equivalent counting problem for outgoing radiation
states. The lifetime of a black hole of mass $M$ in 3+1 dimensional
spacetime is
\begin{equation}
D\sim M^{3}.\label{eq:lifetime}
\end{equation}
The entropy of the black hole is correctly computed if we simply calculate
the out-of-equilibrium entropy of a quasi-thermal gas of photons,
imagining it to be released outward into a region of radius $D.$
The energy of the gas must match the energy of the black hole
\begin{equation}
M\sim T^{4}V\delta\Omega,\label{eq:energy}
\end{equation}
where $T$ is the temperature of the gas, and $\delta\Omega$ is the
solid angle subtended by the past projection of the black hole horizon.
In this case we have
\begin{equation}
\delta\Omega\sim\frac{M^{2}}{D^{2}}\sim\frac{1}{M^{4}}.\label{eq:solidangle}
\end{equation}
Substituting in we then find that 
\begin{equation}
T\sim1/M,\label{eq:temp}
\end{equation}
and the entropy of the gas of radiation 
\begin{equation}
S\sim E/T\sim M^{2}.\label{eq:entropy}
\end{equation}
A more precise calculation would take into account that the temperature
of the gas changes as the black hole evaporates, modulating the energy
density of the outgoing radiation.

We note that if the black hole energy changes by $\delta M$, then
the entropy of the black hole changes by 
\begin{equation}
\delta S=\frac{\delta M}{T_{H}},\label{eq:bhentropy}
\end{equation}
where $T_{H}$ is the Hawking temperature for a mass $M$ black hole.
Thus when emitted, the non-equilibrium outgoing radiation will carry
off this entropy together with any extra contribution arising from
irreversibility in accord with the generalized second law. In four-dimensional
spacetime, the entropy of this quasi-thermal radiation, when treated
like black body radiation, becomes \citep{Zurek:1982zz,mukhanov}\footnote{In $d$ spatial dimensions the prefactor is $\frac{(d+1)}{d}$.}
\begin{equation}
\delta S=\frac{4}{3}\frac{\delta M}{T}.\label{eq:radentropy}
\end{equation}
The extra entropy production in \eqref{eq:radentropy} versus \eqref{eq:bhentropy}
can be viewed as an artifact of the semiclassical approximation, where
the emitted radiation contains no mutual correlations \citep{Zurek:1982zz}.
In \citep{mukhanov} there is an attempt to match this equilibrium
entropy \eqref{eq:radentropy} with \eqref{eq:bhentropy} by adjusting
the temperature $T>T_{H}$. From our perspective, this is unnecessary
since the time reverse of the emitted black hole radiation will not
obey \eqref{eq:radentropy} but rather \eqref{eq:bhentropy} in the
context of unitary evolution. If we consider the full set of ingoing
states obeying \eqref{eq:radentropy}, the important point is that
we obtain a overcomplete set of states for forming small black holes.
We expect the states accounting for the discrepancy between \eqref{eq:bhentropy}
and \eqref{eq:radentropy} will not form black holes, and instead
scatter back out on a much shorter timescale when quantum effects
are accounted for. 

In effect, the Mukhanov conditions amount to requiring the outgoing
states are selected from an ensemble of pure states with energy $E=M$
in a region of finite size $D\sim M^{3}$, with the additional solid
angle constraint \eqref{eq:solidangle}, and the momentum to be predominantly
in the radial direction.\footnote{The region under consideration can be extended to sizes larger than
$D$ but in this case the cloud of outgoing Hawking radiation will
occupy an outgoing shell of width $D$ with vacuum outside and inside
the shell. The radiation is free streaming so the entropy does not
change.} Such constraints are straightforward to implement using wavepackets
of finite width to simultaneously implement the spatial and momentum
constraints. These quasi-local operators can be constructed directly
in holographic theories (assuming the matching between gravitational
and holographic quantities is precise far from the black hole). For
example, in the case of AdS/CFT the methods of HKLL \citep{Hamilton:2005ju,Hamilton:2006az}
can be applied provided $D\gg l_{Pl}$ where $l_{Pl}$ is the Planck
length. The Mukhanov conditions will be useful provided the AdS radius
of curvature satisfies $R_{AdS}\gg D.$ 

More specifically, HKLL construct bulk operators as smeared integrals
of CFT primaries over compact spacetime regions on the boundary. The
resulting expressions are covariant under the conformal group. If
one introduces a time-slicing on the boundary, these smeared integrals
can be further localized to a narrow range of time on the boundary,
by using free boundary propagation. This is sufficient to build a
boundary state, corresponding to the outgoing Mukhanov state. The
time reverse of an outgoing Mukhanov state provides an initial state
for black hole formation and the procedure may be repeated for this
state and arranged so there is no temporal overlap with the final
state operator insertions. Our construction thus provides a map for
both in- and out-going bulk states, that only have support away from
the black hole region and are well described by the semiclassical
gravitational theory, into corresponding states in the boundary theory
which undergo manifestly unitary time evolution. The bulk states can
also be evolved in the semiclassical theory and this bulk evolution
will eventually diverge from unitary evolution generated by the boundary
theory. In the following, we will use the map from the bulk to the
boundary and methods from quantum chaos theory to estimate the size
of corrections to the semiclassical approximation for appropriately
defined observables.

\section{Black Hole Eigenstate Thermalization\label{sec:Quantum-Chaos}}

The eigenstate thermalization hypothesis \citep{srednicki} explains
how an isolated quantum system, which is prepared in a far-from-equilibrium
initial state, can evolve to a state that is hard to distinguish from
thermal equilibrium. The hypothesis, which is expected to hold for
quantum systems whose classical limit exhibits dynamical chaos, originates
from the work of Berry \citep{Berry_1977} and was further developed
by Deutsch \citep{deutsch} and Srednicki \citep{srednicki,Srednicki_1996}.
It has been validated in various model systems but not proven rigorously. 

Analogous behavior arises in gravitational physics when non-thermal
matter undergoes gravitational collapse to a black hole that exhibits
thermal properties. We propose that in a gravitational theory with
a holographic dual this is more than an analogy and that eigenstate
thermalization can explain the validity of semiclassical gravity for
observables that involve a relatively small number of fields. In this
case, an isolated black hole formed by gravitational collapse is described
by a state in the dual field theory that lives in a Hilbert subspace
of dimension $e^{S(E)}$ where $E$ is the energy of the black hole
and $S$ is the Bekenstein-Hawking entropy. The eigenstate thermalization
hypothesis can be expressed as follows \citep{Srednicki_1996}: For
any observable $A$ which is a smooth function of the classical phase
space coordinates, we assume the matrix elements between energy eigenstates
will take the form 

\begin{equation}
A_{\alpha\beta}=\mathcal{A}(E_{\alpha})\delta_{\alpha\beta}+e^{-S\left(\left(E_{\alpha}+E_{\beta}\right)/2\right)/2}R_{\alpha\beta},\label{eq:eth}
\end{equation}
where $\mathcal{A}(E_{\alpha})$ is a smooth function of the energy
and $R_{\alpha\beta}$ is a random matrix, whose matrix elements are
drawn from a Gaussian distribution with variance of order one. There
is a further assumption here that the number of choices for observables
$A$ is not exponential in the system size. The $e^{-S/2}$ factor
in front of the off-diagonal matrix elements can be understood at
a qualitative level by observing that for generic $A$ the matrix
elements of $A^{2}$ should also satisfy \eqref{eq:eth}. To see this,
one inserts \eqref{eq:eth} twice on the right hand side of $(A^{2})_{\alpha\beta}=\sum_{\gamma}A_{\alpha\gamma}A_{\gamma\beta}$
and carries out the sum over $\gamma$. The exponentially large number
of terms in the sum combined with elementary properties of random
matrices gives an estimate that precisely offsets the suppression
from the $e^{-S/2}$ prefactors that accompany the off-diagonal matrix
elements \citep{Srednicki_1996}. Note that it is only necessary to
assume \eqref{eq:eth} holds for the Hilbert subspace relevant for
an isolated black hole, together with its formation and evaporation
products. It need not hold for the entire Hilbert space of the complete
quantum theory. 

Now let us consider eigenstate thermalization in the context of black
hole evolution. We have in mind an isolated black hole, formed from
an initial pure state $|\psi\rangle$ in gravitational collapse that
is reasonably well-localized in time, and then allowed to evaporate
down to nothing without further disturbing it. An initial state of
this form can always be expanded on the basis of states provided by
the holographic construction in Section \eqref{sec:Black-hole-evolution}
but we will also assume that $|\psi\rangle$ is well-described by
semiclassical evolution in the asymptotic region. We then want to
address the question of how long and to what extent the semiclassical
evolution remains faithful to the holographic evolution of the state
forward in time, both outside and inside the black hole. Of course,
it is possible to choose initial states for which the semiclassical
approximation fails from the outset but then our geometric description
also fails from the outset. A simple example would be a superposition
of equal mass states localized in different points in space \citep{Wald:1984rg}.
The holographic theory nevertheless provides a complete description
of the time evolution of such states.

The matrix elements in \eqref{eq:eth} are between energy eigenstates
$|\alpha\rangle$ whose detailed structure depends on the details
of the holographic theory. It is worth noting that the energy eigenstates
themselves will not have a simple geometric/semiclassical description.
Rather they are stationary states formed by superposition of ingoing
and outgoing Mukhanov states for black hole formation at all different
times. The initial state $|\psi\rangle=\sum_{\alpha}C_{\alpha}|\alpha\rangle$
has an energy width given by 
\begin{equation}
\Delta_{\psi}E=\left(\sum_{\alpha}|C_{\alpha}|^{2}\left(E_{\alpha}-\left\langle E\right\rangle _{\psi}\right)^{2}\right)^{1/2},\label{eq:energywidth}
\end{equation}
which has a natural value of order $1/E^{3}$ due to the finite black
hole lifetime. This is a narrow resonance but still $\Delta E$ is
parametrically larger than $e^{-S(E)}$ and therefore expectation
values can self-average over a large number of order $e^{S(E)}$ states.\footnote{Even with finite $\Delta E$ one can build states that do not self-average
over many energy eigenstates. However such states will not satisfy
the semiclassical approximation in the asymptotic region.} We also require the observable $A$ in \eqref{eq:eth} to be sufficiently
smooth across the range of energies under consideration. The precise
criterion is considered in \citep{Srednicki_1996} with the conclusion
\begin{equation}
\left(\Delta E\right)^{2}\ll\left|\frac{A\left(\left\langle E\right\rangle \right)}{A''\left(\left\langle E\right\rangle \right)}\right|.\label{eq:markcondition}
\end{equation}
This criterion is easily satisfied for a large class of observables
in a black hole background where $\Delta E\sim1/E^{3}$ and does not
represent a severe restriction on the local operators available for
semiclassical physics. In fact, eigenstate thermalization applies
for states that are as broad as $\Delta E\lesssim O(1)$ but such
a broad state would be short lived compared to the macroscopic black
holes that are of primary interest here. 

Now let us consider two different normalized pure states, $|\psi\rangle$
and $|\psi'\rangle$ satisfying the above criteria,

\begin{equation}
|\psi\rangle=\sum_{\alpha}C_{\alpha}|\alpha\rangle,\qquad|\psi'\rangle=\sum_{\alpha}C'_{\alpha}|\alpha\rangle,\label{eq:purestates}
\end{equation}
and compute the following quantity, representing the fluctuation of
the expectation value of $A$ between the two states, 
\begin{equation}
\delta A=\left\langle \psi|A|\psi\right\rangle -\left\langle \psi'|A|\psi'\right\rangle =\sum_{\alpha\beta}D_{\alpha\beta}\,e^{i(E_{\alpha}-E_{\beta})t/\hbar}A_{\alpha\beta},\label{eq:achange}
\end{equation}
where $D_{\alpha\beta}=C_{\alpha}^{*}C_{\beta}-C_{\alpha}^{\prime*}C'_{\beta}$
and normalization implies $\sum_{\alpha}D_{\alpha\alpha}=0$. We can
estimate $\delta A$ by using \eqref{eq:eth} for the matrix elements
in the energy eigenstate basis,
\begin{equation}
\delta A=\sum_{\alpha\beta}D_{\alpha\beta}\left(\mathcal{A}(E_{\alpha})\delta_{\alpha\beta}+e^{-S\left(\left(E_{\alpha}+E_{\beta}\right)/2\right)/2}R_{\alpha\beta}\right)e^{i(E_{\alpha}-E_{\beta})t/\hbar}.\label{eq:deltaa}
\end{equation}
Consider the first term, arising from the diagonal contributions,
which will be time independent. For simplicity, let us define $\left\langle E\right\rangle _{\psi}=\left\langle E\right\rangle _{\psi'}$
but we will allow that $\Delta_{\psi'}E=c\Delta_{\psi}E$ with $c=\mathcal{O}(1)$.
Then the leading portion of the first term will be
\begin{equation}
\delta A_{\mathrm{diag}}=\frac{1}{2}(1-c^{2})\left(\Delta_{\psi}E\right)^{2}A''\left(\left\langle E\right\rangle _{\psi}\right),\label{eq:diag}
\end{equation}
and then it directly follows from the criteria \eqref{eq:markcondition}
constraining the width of the band of energy eigenstates that $\vert\delta A_{\mathrm{diag}}\vert\ll\vert A\left(\left\langle E\right\rangle \right)\vert$.
The diagonal corrections depend on the energy profiles of the states
$|\psi\rangle$ and $|\psi'\rangle$ through the slowly varying function
$\mathcal{A}(\left\langle E\right\rangle _{\psi})$ but are otherwise
insensitive to the detailed properties of the microstates. It is natural
to assume that the semiclassical approximation is capable of correctly
computing this contribution, since it should correctly compute quantities,
including finite size corrections, that depend on the expectation
value of the energy. The RST model in two dimensional spacetime provides
a detailed example where such finite size corrections are explicitly
computable \citep{Russo:1992ax}. We also note that the finite size
corrections emerging from 2d CFT show a similar behavior \citep{Datta:2019jeo}.

The off-diagonal corrections, however, depend on the random matrix
$R_{\alpha\beta}$ which depends sensitively on the choice of state.
The semiclassical approximation will not capture those effects correctly.
The size of the off-diagonal contribution to $\delta A$ in \eqref{eq:deltaa}
is governed by the random fluctuations of $R_{\alpha\beta}$ and for
typical microstates $|\psi\rangle$ and $|\psi'\rangle$ one obtains

\begin{equation}
|\delta A_{\mathrm{off-diag}}|\sim e^{-S\left(\left\langle E\right\rangle \right)/2}.\label{eq:offdiagval}
\end{equation}
The answer for the off-diagonal terms in \eqref{eq:deltaa} is time
dependent but at $t=0$ all the phase factors will be equal to one.
On timescales of order the Heisenberg time $t\sim e^{S}$ we expect
this expectation value to recur. 

The estimate in \eqref{eq:offdiagval} will hold for most microstates
but there are special states for which the off-diagonal contribution
is considerably larger. The maximal possible value will be realized
when either $|\psi\rangle$ or $|\psi'\rangle$ happens to be close
to the eigenstate of the matrix $R_{\alpha\beta}$ that belongs to
its largest eigenvalue. It then follows from the Wigner semi-circle
law \citep{Wigner} that $|\delta A_{\mathrm{off-diag}}|_{\mathrm{max}}\sim\mathcal{O}(1)$
but the probability for a randomly chosen microstate to be close to
such an eigenstate is extremely small based on a Haar measure for
$U(e^{S})$. Let us constrain $|C_{\alpha}-\tilde{C}_{\alpha}|<\epsilon\ll1$,
where $\tilde{C}_{\alpha}$ specifies the eigenstate of $R_{\alpha\beta}$
in question. Then the likelihood of picking the state would be of
order $\epsilon^{e^{S}}$, which is negligible. We will therefore
use \eqref{eq:offdiagval} as an estimate of the off-diagonal term
for the states of interest. 

The exponentially small off-diagonal corrections to observables computed
using a semiclassical approximation are impossible to detect unless
one projects onto states that are very close to energy eigenstates.
To probe how the semiclassical approximation would break down outside
the black hole in an operationally well-defined way, we are led to
consider initial and final states for which the off-diagonal corrections
in \eqref{eq:offdiagval} are comparable to the diagonal corrections
coming from \eqref{eq:diag}. To measure violations of the semiclassical
approximation outside the black hole, one must in effect tune $\Delta E\lesssim e^{-S(E)/4}$
which requires enormously long timescales in the asymptotic region
and presumably requires extremely large measuring apparatus and repeated
measurements on identically prepared systems \citep{Harlow:2013tf}.
As noted above, an isolated black hole formed in gravitational collapse
does not have such sharply tuned energy width and we can therefore
expect the semiclassical approximation to be valid for the computation
of expectation values of local operators throughout the geometry outside
the black hole,\footnote{We will turn our attention to observables for infalling observers
in Section \eqref{sec:Infalling-Observables} below.} up to corrections of order $e^{-S(E)/2}$. 

There are, however, key aspects of the physics of black hole evolution
where the semiclassical approximation fails badly. This includes,
for instance, transition amplitudes between in- and out-states. Consider
an ingoing state $|\psi\rangle$ that is well-described by the semiclassical
theory in the asymptotic region, along with a different outgoing state
$|\psi'\rangle$ that also has a good semiclassical description and
compute a matrix element of local operators,
\begin{equation}
\left\langle O(x)O(y)\right\rangle _{\psi\psi'}=\left\langle \psi'|O(x)O(y)|\psi\right\rangle .\label{eq:twopoint}
\end{equation}
From the point of view of the semiclassical approximation, we could
compute this correlator in two ways. First, begin with the initial
state $|\psi\rangle$, evolve it forward in time, producing an outgoing
thermal flux of radiation in accord with Hawking's original computation
\citep{Hawking:1975vcx}. The correlator can then be evaluated as
an expectation value in this semiclassical background.\footnote{The RST model \citep{Russo:1992ax} provides an example where back-reaction
can be fully incorporated at the semiclassical level to justify this
assumption.} On the other hand, we can also begin with the bra-state $\langle\psi'|$
and evolve that backward in time according to the same set of rules,
producing a thermal incoming flux of radiation in accord with the
time reverse of Hawking's calculation. The states $|\psi\rangle$
and $|\psi'\rangle$ can be any states satisfying the Mukhanov conditions,
which allows a wide variation in the energy fluxes, subject to the
average null energy condition. In general, the answers obtained by
the two semiclassical computations will disagree by relative errors
of $\mathcal{O}(1)$ throughout the geometry. In the asymptotic region,
the predicted answers will disagree by relative errors of $\mathcal{O}(1)$
simply because the expectation value of the incoming flux associated
with the state $|\psi\rangle$, which is the input to the semiclassical
solution of the gravitational equations, will differ by corrections
of relative $\mathcal{O}(1)$ with the back-evolved prediction for
the expectation value of the incoming flux associated with the final
state $|\psi'\rangle$, and vice versa. Likewise, near the apparent
horizon the answers will disagree by errors of $\mathcal{O}(1)$.
We conclude that even if $|\psi\rangle$ and $|\psi'\rangle$ lead
to well-behaved semiclassical geometries in the past/future asymptotic
region respectively, the semiclassical approximation fails when computing
the correlator \eqref{eq:twopoint}. This is of course just restating
the information problem, which amounts in this context to the failure
of semiclassical physics to reproduce the unitary S matrix of the
holographic dual theory.

Another example where semiclassical physics might be expected to fail
is in computing non-local quantities such as entanglement entropy
between the black hole and the outgoing radiation. Since we assume
the black hole involves a finite dimensional Hilbert subspace under
unitary evolution, and follows the rules of quantum mechanics, Page's
calculation of entanglement entropy immediately follows \citep{Page:1993df}.
In particular, the entanglement entropy will follow a so-called Page
curve and fall to zero at late times for any incoming pure state.
In an interesting development \citep{Almheiri:2019psf,Penington:2019npb},
a Page curve was obtained using semiclassical methods for a large
AdS black hole coupled to an external conformal field theory reservoir
by adapting a geometric prescription for calculating generalized entropy
\citep{Ryu:2006bv,Hubeny:2007xt,Faulkner:2013ana,Engelhardt:2014gca}.
The fact that a fine-grained quantity like entanglement entropy is
correctly reproduced by a semiclassical prescription is highly non-trivial
but does not on its own provide much insight into the underlying quantum
physics. 

\section{Infalling Observables\label{sec:Infalling-Observables}}

To build observables for infalling observers we begin with observations
made in our previous work \citep{Lowe:2015eba}. Consider a black
hole that has been formed, and then left isolated for a scrambling
time, of order $\beta\log S$ where $\beta$ is the inverse Hawking
temperature. After this scrambling phase \citep{Sekino:2008he}, we
assume we can apply \eqref{eq:eth} to the holographic dual. We presume
the black hole has been formed from a Mukhanov state that is well-described
by the semiclassical approximation (\emph{i.e.} not, for example,
a macroscopic superposition of well-separated black holes). According
to the results of \citep{Lowe:2015eba} an effective field theory
may be set up on a set of timeslices corresponding to a freely infalling
lattice of spatial points moving along timelike geodesics \citep{Corley:1997ef}.
Local operators corresponding to infalling ``labs'' may then be
evolved across the horizon. Moreover the Hamiltonian that evolves
these operators along timelike geodesics may also be constructed in
the same manner. We assume that outside the horizon, the lab states
are well-described by a semiclassical approximation, and do not involve
strong gravitational effects such as the formation of additional black
holes inside the lab. Due to the physical cutoff \citep{Lowe:2015eba},
the time evolution of such operators (and their expectation values)
becomes independent of the details of what was sent in earlier (\emph{i.e.}
more than $\beta\log S$ earlier). Thus from the gravity perspective,
the details of the black hole microstate do not influence the evolution
of these operators along timelike geodesics. It remains to check then
whether this semiclassical evolution remains accurate when computed
using holographic methods.

Outside the horizon of the black hole, in the vicinity of the region
where the Mukhanov states are set up, the lab operators $\phi(x)$
can be mapped precisely into the holographic description. Likewise
the infalling Hamiltonian density acting on this set of operators
may be defined as an operator in the holographic theory. Our philosophy
will be that the expectation values of this family of operators (\emph{i.e.}
including their fluctuations $\left|\phi(x)-\left\langle \phi(x)\right\rangle \right|^{2}$)
provides a complete description of the observables accessible to a
local infalling observer. Consider
\begin{equation}
|\mathrm{lab+black}\,\mathrm{hole}\rangle=\left\{ \phi\right\} |\psi\rangle.\label{eq:labket}
\end{equation}
This state will typically live in a larger Hilbert subspace than the
black hole itself. Here $\left\{ \phi\right\} $ is some set of operators
built of products and sums of the $\phi$'s. The expectation values
that describe the infalling observer will involve matrix elements
with states such as
\begin{equation}
\mathrm{\langle lab+black}\,\mathrm{hole}|=\langle\psi|\left\{ \phi\right\} '.\label{eq:labbra}
\end{equation}
Here $\left\{ \phi\right\} '$ is some different set, describing the
lab at some future time. These matrix elements can then be computed
using the methods described in Section \eqref{sec:Quantum-Chaos}
above, 
\begin{equation}
\langle\psi|\left\{ \phi\right\} '\left\{ \phi\right\} |\psi\rangle=\sum_{\alpha\beta}C_{\alpha}C_{\beta}^{*}\langle\beta|\left\{ \phi\right\} '\left\{ \phi\right\} |\alpha\rangle=\sum_{\alpha\beta}C_{\alpha}C_{\beta}^{*}A_{\alpha\beta}.\label{eq:infallermatrix}
\end{equation}
These expectation values can be computed to an accuracy that depends
on $\Delta E$ for the choice of black hole states, and the entropy
of the black hole. For $\Delta E\sim O(1/E^{3})$, we expect they
will be computed to a good approximation by semiclassical methods,
as described above.

Now if we have too many independent operators, one can always build
linear combinations for which the above statement is no longer true.
This bound becomes sharpest when we minimize $\Delta E$ to the point
where the corrections are not captured by the semiclassical approximation,
\emph{i.e.} $\Delta E\sim e^{-S/4}$. For an infaller to detect such
corrections, they need access to an enormous number of order $\mathcal{N}_{lab}\sim e^{S/2}$
operators.

The number of operators an infalling lab has access to can be estimated
in a variety of ways. One would be to simply use a holographic bound
and assume that if the infalling space of states lives in a Hilbert
subspace of dimension $e^{S_{BH}(E_{lab})}$ then the dimension of
the space of operators would be of order $e^{2S_{BH}(E_{lab})}$.
For $E_{lab}\ll M$ this number is far less than $e^{S(M)/2}$.

An alternative would be to use a spatial lattice cutoff as in \citep{Lowe:2015eba}
and combine that with a lattice cutoff in the time direction. With
physically reasonable choices for these cutoffs, it is again easy
to obtain an estimate that is far less than $e^{S(M)/2}$ assuming
the energy of the lab is much smaller than the energy of the black
hole. Again we conclude that the family of matrix elements \eqref{eq:infallermatrix}
can be computed to high accuracy by the semiclassical approximation.
Once this is established we then have a self-consistent approximation
which allows us to push these operators forward along their timelike
geodesics into the interior of the black hole, using evolution with
respect to the infalling time translation operator and compute their
expectation values to the precision stated. 

The semiclassical evolution will eventually deviate significantly
from the unitary evolution of the corresponding states in the holographic
dual theory. The time scale on which this occurs was estimated in
a simplified holographic model in \citep{Lowe:2016mhi,Lowe:2017ehz}
and found to match the black hole scrambling time. Interestingly,
this is also the maximum time an infalling observers can avoid the
black hole singularity in the infalling lattice model of \citep{Corley:1997ef,Lowe:2015eba}.
The holographic model resolves the black hole singularity in the sense
that approaching the region of strong curvature coincides with the
breakdown of the semiclassical approximation for infalling observables
but the holographic evolution remains well defined. 

\section{Conclusions\label{sec:Conclusions}}

Even when the semiclassical approximation holds in the asymptotic
region, it breaks down with corrections of $\mathcal{O}(1)$ for transition
amplitudes. To make an analogy with particle physics, we are pointing
out that rare decays of the black hole can only be explained by supplementing
the semiclassical picture of spacetime near the horizon of the black
hole by new quantum effects that are of the same order of magnitude
as the classical terms. On the other hand, for almost all local observables
of physical interest, self-averaging over of order $e^{S}$ energy
eigenstates occurs. This renders the semiclassical approximation (with
finite size corrections included) accurate up to corrections of order
$e^{-S/2}$. Until we are able to conduct repeated experiments involving
black hole formation and evaporation in the lab, it seems difficult
to discriminate between different theories of quantum gravity that
obey these rules. An exception to this would be repeated observations
of the end point of Hawking evaporation where the effective $S$ is
of $\mathcal{O}(1)$ and repeated observations of even local observables
will depend in detail on the structure of the underlying theory of
quantum gravity.

The experience of infalling observers can likewise be recast as correlators
of local observables initially located outside the horizon. As explained
in \citep{Lowe:2015eba,Lowe:2016mhi,Lowe:2017ehz}, these observables
can be evolved along timelike geodesics and their time evolution can
be evaluated. Given the constraints on the types of projection operators
that can be built from such observables, the self-averaging over $e^{S}$
states will still be in play, and we can in principle evaluate such
observables to an accuracy of order $e^{-S/2}$ using a semiclassical
approximation until they enter the high curvature region well-inside
the horizon.

These results offer support to the black hole complementarity paradigm
for solving the black hole information problem \citep{Susskind:1993if}.
Information propagating across the horizon is well-described by semiclassical
evolution toward the curvature singularity. At the same time, expectation
values of local operators outside the horizon receive corrections
of order $e^{-S/2}$ which permits unitary evolution to an outgoing
cloud of Hawking radiation. We have identified criteria which lead
to accurate semiclassical evolution of states. Physically interesting
states almost always satisfy these criteria. Conversely precise criteria
can be given for states which violate the semiclassical approximation
and which cannot be interpreted geometrically. 
\begin{acknowledgments}
Work supported in part by DOE grant de-sc0010010 Task A, Icelandic
Research Fund grant 228952-051, and a grant from the University of
Iceland Research Fund. 
\end{acknowledgments}

\bibliographystyle{utphys}
\bibliography{coherent}

\end{document}